# Salinity and sodicity can cause hysteresis in soil hydraulic conductivity

Taiwo Adeyemo[1], Isaac Kramer[1], Guy J. Levy[2], Yair Mau[1*]

[1] Institute of Environmental Sciences, The Robert H. Smith Faculty of Agriculture, Food and Environment, The Hebrew University of Jerusalem, Rehovot 76100 Israel

[2] Institute of Soil, Water and Environmental Sciences, Agricultural Research Organization - Volcani Institute, Israel

* Corresponding author

E-mail addresses: taiwo.adeyemo@mail.huji.ac.il (T. Adeyemo), isaac.kramer@mail.huji.ac.il (I. Kramer), vwguy@volcani.agri.gov.il (G.J. Levy), yair.mau@mail.huji.ac.il (Y. Mau).

**Abstract**

Soil salinization and sodification can cause detrimental effects to soil structure, with important implications to irrigated agriculture. Knowledge of the extent to which degradation in soil structure due to salinization and sodification is reversible is still lacking, however. The objective of our study is to examine the effects of the history of solute composition on the degree of reversibility in saturated hydraulic conductivity ($K_{sat}$). We systematically investigate the effects of salinity (electrolyte concentration) and sodicity (sodium adsorption ratio) on $K_{sat}$, for three soils of varying clay content. The soil column experiments yielded hysteresis graphs, in which $K_{sat}$ does not go back to its original values after initial decay. We developed indices to quantify the degree of $K_{sat}$ degradation and reversibility, and found that contrary to our expectations, high susceptibility to degradation does not always correlate with low capability of rehabilitation. Measurements of soil swelling helped us discern when degradation was mainly caused by swelling or clay dispersion. Our findings underscore the need of understanding hysteresis caused by salinity and sodicity, aiming at a better management of agricultural soils for sustainable long-term use.

**Highlights**:

- Hydraulic conductivity does not completely rehabilitate after initial degradation
- Soils with higher clay content are more susceptible to degradation
- High degradation risk does not necessarily correlate with limited rehabilitation
- The degree of rehabilitation of a soil depends on its texture and sodicity levels

**1. Introduction**

Problems associated with sodic soils have long been recognized (e.g., special issue of the Australian Journal of Soil Research, [vol. 31 no. 6, 1993]; Sumner and Naidu, 1998; Levy 2012) and are, unfortunately, expected to increase in the future. Yet, the ever increasing need to provide food to an expanding worldwide population, coupled with the increasing demand for good quality water from urban and industrial sectors, result in poorer quality water and soils (i.e., more sodic and saline) being used for food production. Consequently, improving our understanding of the combined impact of sodicity and salinity on agriculture and the soil ecosystem is essential in order to properly manage sodic soils for crop production and maintain sustainable agriculture in many parts of the world. A recent review by Hopmans et al. (2021) points out, however, that the current knowledge and understanding of the effects of sodicity and salinity on soil physical and hydraulic properties is still limited.

Two important parameters are generally used to define soil sodicity (Levy 2012). The first is the exchangeable sodium percentage (ESP, dimensionless) which describes the fraction of adsorbed sodium ($Na_x$) from the cation exchange capacity (CEC) of the soil, and is defined as:

$$ESP = 100 \frac{Na_x}{CEC}. \qquad [1]$$

The second parameter indicates the sodicity level of the soil solution. It is termed sodium adsorption ratio (SAR) and is defined as:

$$SAR = \frac{[Na]}{\sqrt{\frac{[Ca]+[Mg]}{2}}}, \qquad [2]$$

where the square brackets indicate cation concentration in mmol$_c$ L$^{-1}$. Thus, SAR has units of (mmol$_c$ L$^{-1}$)$^{1/2}$.

To date there is no widely accepted definition of a sodic soil. It has been suggested that ESP of 15 should separate sodic from non-sodic soils (USSL Staff, 1954; McNeal and Colman, 1996), yet it was added that this limit must be regarded as somewhat arbitrary and tentative, and that the electrolyte concentration in the percolating solution must exceed 3 mmol$_c$ L$^{-1}$ in order for this ESP level to hold true. Later on, Greene et al. (1978) proposed that the threshold value separating sodic from nonsodic soils should depend on soil texture; they proposed ESP values of 10 and 20 for fine and coarse textured soils, respectively. More recent hydraulic conductivity (HC) studies by Mace and Amrhein (2001) and Levy et al. (2005) for calcareous semi-arid soils indicated that soils with ESP > 5 exposed to solutions with electrolyte concentration (C) of 0.7 mmol$_c$ L$^{-1}$, should be considered as Na-affected (sodic) soils. The abovementioned discussion also highlights the importance of the interrelationship between the SAR and the C of the soil solution in dictating whether sodic or non-sodic behavior will be observed (see detailed discussion later on).

Colloidal clay determines much of the physical behavior of soils because of its large specific surface area and charge, which makes it very reactive in physicochemical processes and in particular swelling and dispersion (e.g., Levy, 2012). These two processes determine, to a large extent, soil microstructure (Bennett et al., 2019) and thus many of its hydraulic properties including saturated hydraulic conductivity. A comprehensive discussion of the processes involved in clay dispersion and swelling are presented by Goldberg et al. (2012).

The realization that soil HC depends on both the SAR and the C of the percolating solution, led Quirk and Schofield (1955) to develop the "threshold concentration" concept, which was defined as the C required to prevent a decrease greater than 25% in the HC for a given soil ESP or SAR of the percolating solution. This concept explains observations of decreased HC even in Ca-soils and soils of low ESP, exposed to water of low salinity such as rain or snow water (Emerson and Chi, 1977). The observed decrease in HC was ascribed to salt concentration in the soil solution being insufficient to prevent clay swelling and dispersion (Shainberg and Letey, 1984). The importance and dominance of clay swelling and dispersion in controlling the HC has been demonstrated in numerous studies (e.g., McNeal et al., 1966; Cass and Sumner, 1974; Radcliffe et al., 1987; Dikinya et al., 2007; Bauder et al., 2008). Moreover, the effect of swelling on HC is deemed reversible, decreasing the C of the soil solution followed by an increase in the C leads to a decrease and subsequent increase in soil HC (Shainberg and Letey, 1984, van Dijk, et al., 2016). Conversely, when clay dispersion occurs, the dispersed clay particles can move through the soil profile and may even cause a complete blockage of conducting pores, and hence yield irreversible changes in the HC (Shainberg and Letey, 1984).

Although numerous studies have explored the relationship between the electrolyte concentration, SAR, and their effect on soil HC, little research has been devoted to looking at the extent to which changes in the HC of the soil are reversible when subjected to the influence of salinity and sodicity. McNeal & Coleman (1966) reported that the decrease in saturated HC with a decrease in C and increasing SAR in montmorillonitic soils was largely irreversible even when high concentration calcium solutions were used; only soils containing more than 10% montmorillonite showed some reversibility of the saturated HC. An increase in HC after an initial decrease, following leaching with solutions of known concentration and SAR, was noted only after the soil was dried and repacked (Dane & Klute, 1977). Recently, van der Zee et al. (2014) stated that a decrease of saturated HC imposed by sodicity is practically irreversible at a time scale of decades.

The need to maintain sustainable irrigated agriculture in semi-arid and arid regions worldwide cannot be overstated. There is a large body of information on the combined adverse impact of sodicity and salinity on the degradation of soil HC. Yet, there are no systematic studies on possible hysteresis in saturated HC inflicted by salinity and sodicity. Hence, there are numerous questions that are left unanswered, such as: are there salinity and sodicity thresholds beyond which rehabilitation of HC is not possible? What role does soil texture play in this phenomenon? In order to fill, at least some of the gaps in our knowledge, we proposed to test the following **hypotheses**: (i) There are regions in the parameter space (C, SAR) where the degradation in saturated HC is reversible and there are others where it is irreversible or only partially reversible, and (ii) soils of different textures have different contour of regions in the parameter space where the HC hysteresis varies from total reversibility to total irreversibility. Consequently, the **specific objective** of our study was to examine the degree of reversibility in saturated HC through a systematic investigation of the effects of the history of solute composition (C, SAR) on the saturated HC of three soils varying in texture.

## 2. Materials and methods

*2.1 Soils*

Three different non-sodic soils from Israel, varying in texture, were used for the experiment: a sandy loam from the Northern Negev, a sandy clay loam from Kiryat Gat, and a clay from Revadim. Soil samples from each soil type were collected from a depth >50 cm to ensure that the samples contained limited amounts of organic matter, so that it would not markedly affect our study. The soil samples were brought to the laboratory, air-dried, passed through a 2 mm sieve, and stored in a cool place. Selected properties of the soils, determined using standard analytical methods (Klute, 1986, Page et al., 1986), are presented in Table 1.

**Table 1:** Selected properties of the three soil types used.

| Site | Soil type[1] | USDA Classification[2] | Sand (%) | Silt (%) | Clay (%) | CEC[3] $mmol_c\ kg^{-1}$ | ESP[4] | $CaCO_3$ (%) | OC[5] (%) | Hygro-scopic Water (%) |
|---|---|---|---|---|---|---|---|---|---|---|
| Northern Negev | Sandy loam | Typic Haploxeralf | 72 | 18 | 10 | 7.67 | 2.04 | 12.8 | 0.06 | 1.95 |
| Kiryat Gat | Sandy clay loam | Typic Haploxerert | 50 | 25 | 25 | 19.61 | 1.07 | 16 | 0.39 | 3.99 |
| Revadim | Clay | Chromic Haploxerert | 25 | 30 | 45 | 29.88 | 0.70 | 11 | 0.97 | 5.42 |

[1] Soil type based on texture according to the US classification system
[2] Soil classification based on the USDA Soil Taxonomy system
[3] CEC = cation exchange capacity
[4] ESP = exchangeable sodium percentage
[5] OC = organic carbon

*2.2 Solutions studied*

For our experiment we used an array of solutions which comprised five different SAR levels, 2, 5, 10, 20 and 50 $(mmol_c\ L^{-1})^{1/2}$ and four electrolyte concentrations, 5, 20, 80, and 200 $mmol_c\ L^{-1}$. All together twenty solutions (5 SAR times 4 electrolyte concentrations) with the various combinations of electrolyte concentration and SAR were prepared using NaCl and $CaCl_2$ salts. In what follows, SAR units will be omitted for simplicity's sake.

*2.3 Saturated hydraulic conductivity*

Saturated HC was determined in the laboratory using soil columns. Soil columns were prepared by packing 149.8 g of air-dried sieved soil (< 2.0 mm) into Perspex cylinders (5.4 cm internal diameter and 12 cm long) to a dry bulk density of 1.3 g cm$^{-3}$. The bottom of each cylinder was closed with a rubber stopper with a single drilled hole which served as an exit for the leachate. The upper side of the stopper was lined with a geotextile (geotex) fiber that served as a filter for the soil. Another layer of geotex fibre was added to the top of the packed soil.

The soil columns were initially wetted from the bottom with a solution of a given SAR and high concentration of 200 mmol$_c$ L$^{-1}$ through capillary rise (using a Mariotte bottle) at a slow rate to avoid entrapment of air and aggregate slaking. After reaching saturation, the columns were wrapped with aluminum foil to avoid algae proliferation, aimed at reducing any changes in saturated hydraulic conductivity not caused by the treatment studied. Thereafter, flow rate was reversed, and the columns were leached from the top with a given solution using a constant head device (Mariotte bottle) of 34.5 cm.

The leachate was collected continuously over fixed time intervals that depended on the rate of leachate outflow, using tubes. Cumulative flowing water volume (V, mL) was monitored versus time (t, min), to provide data (together with soil column length [L, cm], the hydraulic head [$\Delta H$, cm] and column cross sectional area [A, cm$^2$]) necessary to calculate the saturated HC (K$_{sat}$, cm h$^{-1}$), on the basis of Darcy's law:

$$K_{sat} = \frac{V \cdot L}{A \cdot t \cdot \Delta H} \qquad [3]$$

*2.3 Experimental setup*

The soil columns were leached with solutions of a given fixed SAR and a successively decreasing electrolyte concentration, starting with a concentrated solution of 200 mmol$_c$ L$^{-1}$ and ending with a dilute solution of 5 mmol$_c$ L$^{-1}$. The switch from a concentrated to a more dilute solution took place when the following equilibrium conditions were attained: (i) the flow rate of the leachate volume collected over consecutive intervals was steady, and (ii) the measured EC (an expression for C) of and the Na concentration in the leachate were similar to those in the leaching solution placed in the Mariotte bottle. Switching of the leaching solutions from a concentrated level to a more dilute one in the Mariotte bottle was done carefully to avoid (i) air bubbles entering the system and (ii) sudden changes in the pressure formed in the system, that could adversely affect the $K_{sat}$ measurements during the switch (Moutier et al., 1998).

Along the $K_{sat}$ degradation curve with decreasing C, we selected several points where the trend was reversed to increasing C values. These curves are called reversal curves, and they help us check the system's capability of returning to the original $K_{sat}$ values after degradation has set in. All measurements were carried out in three replicates.

In addition, to determine possible swelling of the soil, the height of the packed soil in each column was measured (i) immediately after dry packing (DP), (ii) after wetting (AW) and (iii) at the end of the leaching experiment (ALE), that is, after completing a whole cycle of leaching from a concentrated (200 $mmol_c$ $L^{-1}$) to more dilute C solutions and back again to the concentrated solution.

*2.4 Degradation and reversibility indices*

In order to quantitatively compare the response of the three soils to changes in salinity and sodicity, we introduce the degradation index and the reversibility index.

The degradation index measures the degree at which a soil's $K_{sat}$ is prone to decay upon a monotonic decrease in C. It ranges from 0 to 1, where 1 means maximal susceptibility to degradation, and 0 denotes no degradation whatsoever.

The reversibility index is drawn from our recent developments in the modelling of $K_{sat}$ hysteresis (Kramer et al., 2021), where the weight functions and reversibility index for each hysteresis diagram are obtained. The weight function fully encapsulates the soil-specific $K_{sat}$ response to changes in salinity and sodicity. From the weight function, we can calculate the reversibility index of each soil by measuring how it responds to $K_{sat}$ degradation; an index of 1 indicates that the soil has complete capability of rehabilitating, while an index of 0 denotes that any degradation is completely irreversible. This index can be useful to quantify differences in $K_{sat}$ behavior between soils of various textures, or even the response of the same soil to different SAR levels.

The calculation of these indices are presented in greater detail in Section 3.3, after we show all the hysteresis curves in Section 3.1. We will show a practical example of how the degradation and reversibility indices are calculated, and we will then compare the indices obtained for all soils and discuss the implications of the results.

*2.5 Data analysis*

Analysis of variance (ANOVA) and Tukey's HSD test ($p < 0.05$) were used to verify significant differences in clay swelling across all three soils resulting from the various SAR and C treatments.

## 3. Results

*3.1 Hysteresis in saturated hydraulic conductivity ($K_{sat}$)*

The curves for $K_{sat}$ vs C for SAR 2 and 5 showed either no hysteresis or even an increase in $K_{sat}$ during the course of decreasing C followed by an additional increase in $K_{sat}$ during the subsequent course of increasing C (Figs. S1 and S2, Supplementary Material). It is evident that the response of $K_{sat}$ to these SAR levels is not typical to sodic soils, therefore, from here onwards we will focus on the results obtained for SAR 10, 20 and 50.

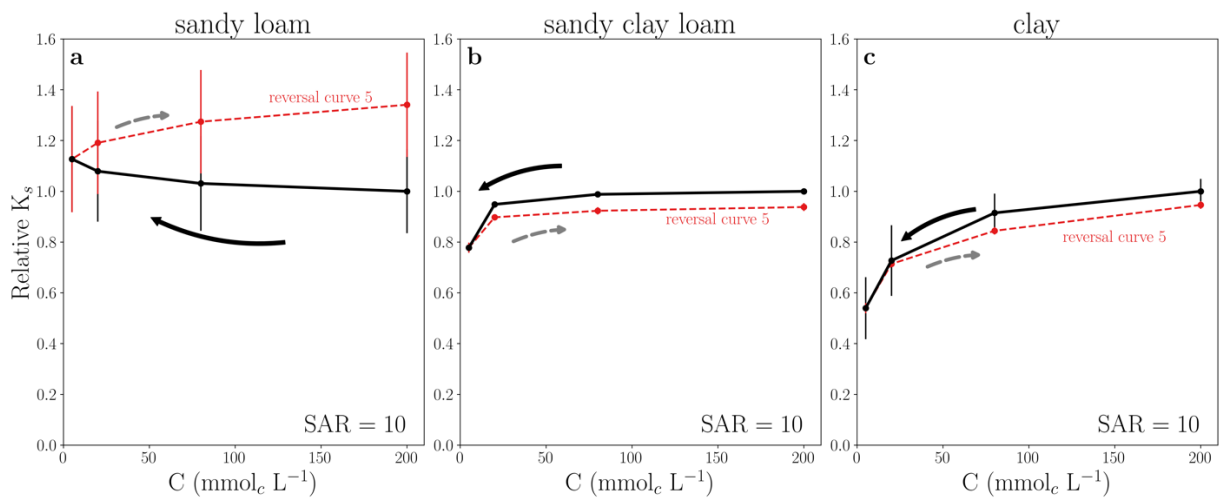

**Figure 1**: Hysteresis curves for the (a) sandy loam, (b) sandy clay loam and (c) clay soils, leached with SAR=10 solutions. Measurements start at the rightmost data point in the solid black curve and follow the arrows. Reversal curves start at the leftmost data point of the dashed lines (5 mmol$_c$ L$^{-1}$) and follow the dashed arrow. Bars indicate standard deviation.

Figure 1 shows the relative $K_{sat}$ response to a decrease and subsequent increase in C for the three soils, when leaching with solution of SAR 10. The relative $K_{sat}$ was calculated by normalizing the measured saturated hydraulic conductivity by the $K_{sat}$ value of the first measurement, namely, the rightmost data point in each solid black curve. The sandy loam $K_{sat}$ was not significantly affected by stepwise decreases in the C of the leaching solution from 200 $mmol_c$ $L^{-1}$ ($K_{sat}$ = 5.5 ± 0.9 cm $h^{-1}$) to a concentration of 5 $mmol_c$ L-1 ($K_{sat}$ = 6.1 ± 1.1 cm $h^{-1}$). Moreover, a subsequent stepwise increase in C back to a concentration of 200 $mmol_c$ $L^{-1}$ led to a gradual increase in $K_{sat}$ to 7.4 ± 1.1 cm $h^{-1}$ (Fig. 1a). This type of hysteresis was contrary to our expectations. In the sandy clay loam and clay soils, a relatively thin hysteresis loop in the $K_{sat}$ values was noted between the course of decreasing C and the subsequent increase in C (Fig. 1b,c). However, $K_{sat}$ in the clay soil seemed more sensitive to the reduction in C, decreasing 46% from 0.43 ± 0.02 to 0.23 ± 0.05 cm $h^{-1}$ with the reduction in C from 200 to 5 $mmol_c$ $L^{-1}$, respectively, compared with the observed milder decrease of 22% in the sandy clay loam (Fig. 1b,c).

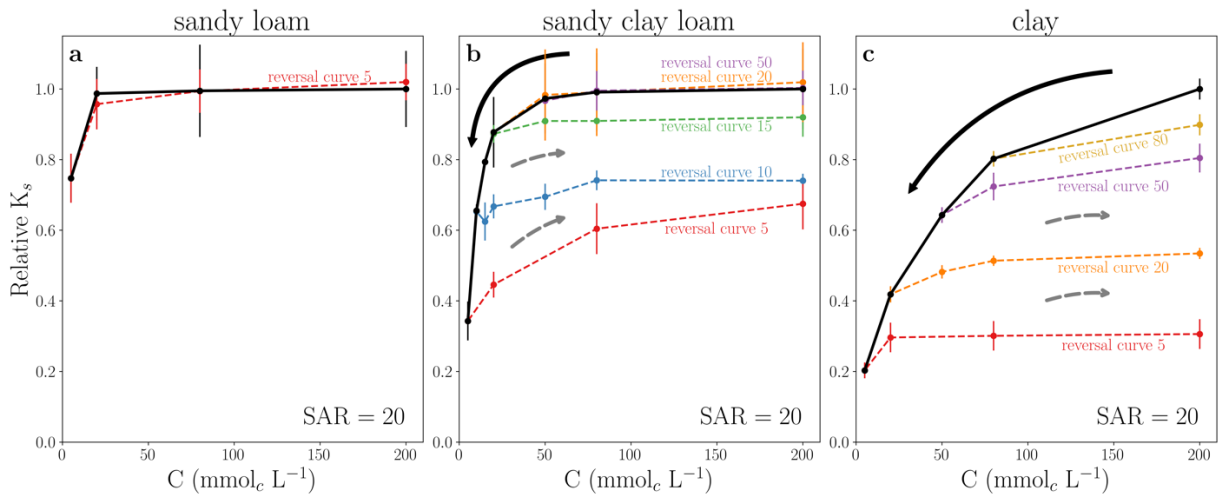

**Figure 2**: Hysteresis curves for the (a) sandy loam, (b) sandy clay loam and (c) clay soils leached with SAR = 20 solutions. Measurements start at the rightmost data point in the solid black curve and follow the solid line arrows. Reversal curves, where the trend in concentration change was shifted from decrease to increase, are marked by colored dashed curves and follow the dashed arrows. The number under each dashed line indicates the concentration at which the reversal curve started. Bars indicate standard deviation.

Leaching the sandy loam with SAR 20 solutions of different concentrations did not yield any hysteresis. In addition, $K_{sat}$ remained unchanged through the leaching except for the leaching with the 5 mmol$_c$ L$^{-1}$ solution, which led to a significant decrease (25%) in $K_{sat}$ (Fig. 2a).

Full hysteresis graphs were noted for the sandy clay loam and the clay soils, following initial $K_{sat}$ degradation with decreasing C (Fig. 2b,c). By 'hysteresis graph' we denote the degradation curve (in black), together with all the reversal curves measured. The sandy clay loam (Fig. 2b) experienced $K_{sat}$ reductions for C values below 50 mmol$_c$ L$^{-1}$. Complete reversibility existed for C dilutions down to 20 mmol$_c$ L$^{-1}$, i.e., the reversal curves for 50 and 20 mmol$_c$ L$^{-1}$ backtracked the degradation curve. Conversely, the reversal curves for 15, 10, and 5 mmol$_c$ L$^{-1}$ could not retrace the degradation curve, thus characterizing hysteresis. For these curves, a partial increase in $K_{sat}$ took place, with diminishing returns as C was increased during the reversal process. It is worth noting that a small concentration difference separates total reversibility in $K_{sat}$ (reversal curve 20), and partial reversibility (reversal curve 15).

The clay soil (Fig. 2c) is more sensitive to degradation, experiencing significant $K_{sat}$ reductions from the onset of concentration decrease. All reversal curves show partial $K_{sat}$ reversibility, with lower degrees of reversibility, as more pronounced degradation took place. This observation implies that the greater the initial decrease in $K_{sat}$, the harder it is to increase it upon reversal in concentration trends.

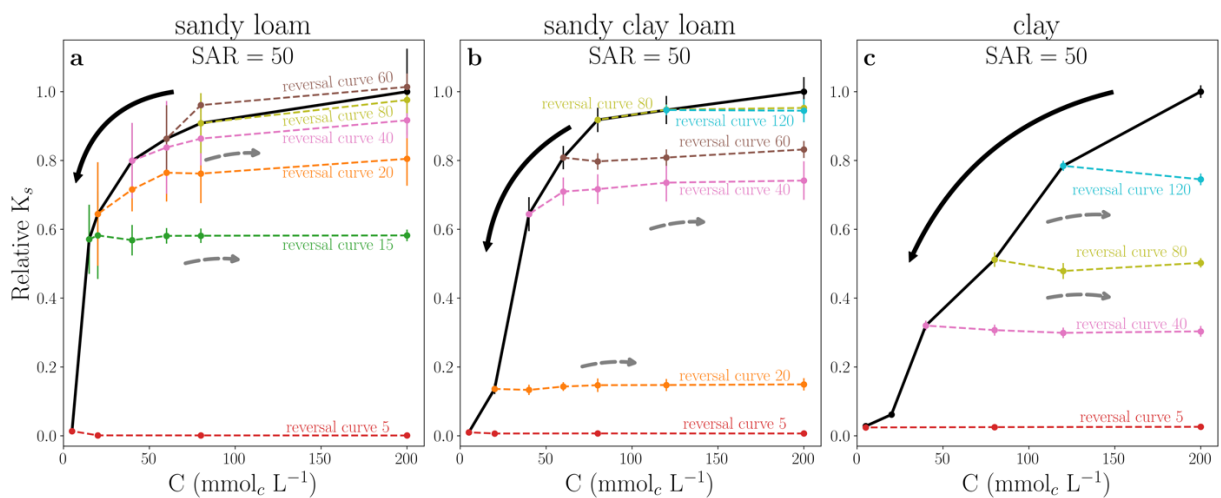

**Figure 3**: Hysteresis curves for the (a) sandy loam, (b) sandy clay loam and (c) clay soils leached with SAR = 50 solutions. Measurements start at the rightmost data point in the solid black curve and follow the solid line arrows. Reversal curves, where the trend in concentration change was shifted from decrease to increase, are marked by colored dashed curves and follow the dashed arrows. The number under each dashed line indicates the concentration at which the reversal curve started. Bars indicate standard deviation.

Leaching the three soils with the SAR 50 solutions resulted in a sharp decrease in $K_{sat}$ during the decrease in solution C, which led to relative $K_{sat}$ values of close to zero, i.e., a near complete blockage of the soil column (reversal curve 5 in all panels of Fig. 3). In the case of the sandy loam and sandy clay loam (Fig. 3a,b) a sharper reduction in $K_{sat}$ took place only once a threshold C level of 40 and 60 mmol$_c$ L$^{-1}$, respectively, was reached. Unlike the two loamy-textured soils, $K_{sat}$ in the clay soil exhibited a nearly linear decrease already from the outset of concentration decrease (Fig. 3c).

We found that the hysteretic behavior in $K_{sat}$ varies widely among these three soils. The sandy loam showed regions of (i) complete $K_{sat}$ reversibility (reversal curves 80 and 60 backtrack the solid black curve), (ii) partial reversibility (reversal curves 40 and 20 go up, but do not reach the initial $K_{sat}$ value), and (iii) complete irreversibility (reversal curves 15 and 5 are horizontal, showing no $K_{sat}$ rehabilitation). Once more, small differences in the degree of $K_{sat}$ degradation can lead to very different results, e.g., reversal curve 20 was able to go back to 80% of the original $K_{sat}$ value, while reversal curve 15 never rehabilitated, staying at 58% of the initial $K_{sat}$.

Both the sandy clay loam and the clay showed either very low $K_{sat}$ reversibility or complete irreversibility (Fig. 3b,c). All the reversal curves are roughly horizontal, indicating no capability of regaining hydraulic conductivity, even for very high concentrations.

*3.2 Soil swelling*

The increase in height of the soils packed in the columns after (1) wetting (i.e., for a C concentration of 200 mmol$_c$ L$^{-1}$, AW), and (2) at the end of the leaching experiment (also for a C concentration of 200 mmol$_c$ L$^{-1}$, ALE), relative to the height of the packed dry soil (5 cm), was used as a proxy for evaluating soil clay swelling under the various SAR solutions studied. The results are shown in Figure 4. No significant increase in soil height, compared with the dry soils, was noted in the columns of the sandy loam at any of the two aforementioned stages (panel a), irrespective of the SAR treatment. In the sandy clay loam (panel b), similar and significant increases in height, relative to the dry soil, were noted in soil columns leached with SAR 2, 5, 10 and 20 for both stages 1 and 2. In the columns leached with SAR 50, the increase in height at stage 2 was significantly greater than that at stage 1. In the clayey soil (panel c), significant differences in the increase in height of the soil in the columns between stage 1 and stage 2 were noted for SAR 10, 20 and 50. For SAR 2 and 5 the increase in height was similar for the two stages but significantly higher than that in the dry soil. Evidently, significant degrees of swelling, that depended on both the SAR of the leaching solution and soil type, took place in the soil columns during the experiment.

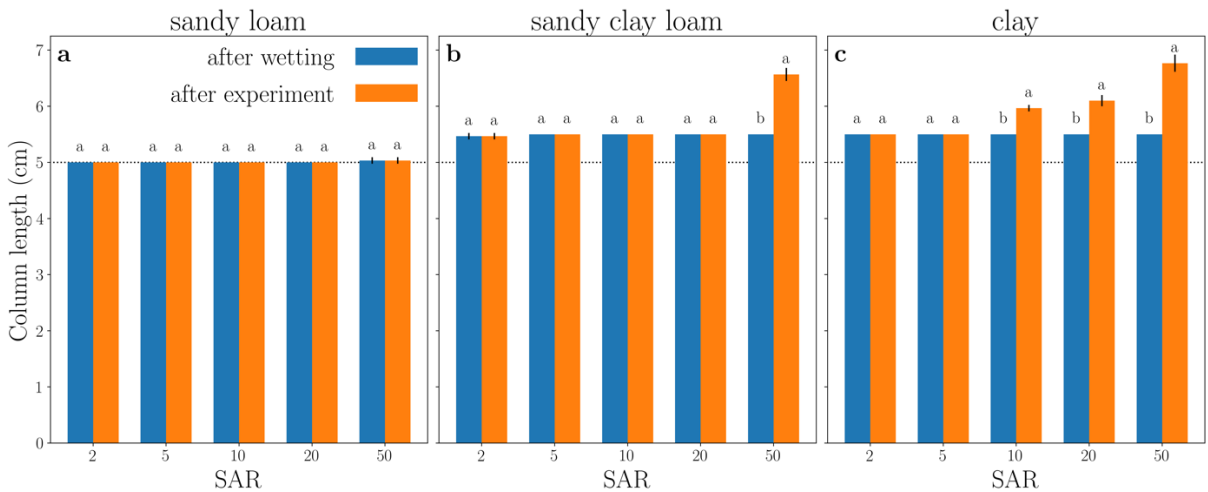

**Figure 4**: Mean swelling response of the (a) sandy loam, (b) sandy clay loam, (c) clay, to the various SAR treatments after wetting (AW) and after the leaching experiment (ALE). Within a given SAR, columns labeled by the same letter do not differ significantly at *p<0.05*.

*3.3 Degradation and reversibility indices*

The degradation and reversibility indices are calculated from the hysteresis curves. Figure 5a shows the hysteresis diagram obtained for the clay soil at SAR 20 (previously seen in Fig. 2c). The solid black curve denotes the degradation curve, namely how $K_{sat}$ responds to decreases in concentration, starting from $C = 200$ mmol$_c$ L$^{-1}$. This same curve is shown in Fig. 5b, where the region below it is hatched with red lines. If the soil were to experience no decrease in saturated hydraulic conductivity, then the curve for $K_{sat}$ would be a horizontal line at relative $K_{sat} = 1.0$. The region below this horizontal line is shaded in gray in Fig. 5b. We define the degradation index (DI) as:

$$\text{DI} = 1 - \frac{\int_{C_{min}}^{C_{max}} k(C) dC}{(C_{max} - C_{min}) k(C_{max})}, \quad [4]$$

where k(C) is the degradation curve in black, and $C_{max}$, $C_{min}$ are the maximum and minimum values of C. The numerator in Eq. (4) is the area hatched with red lines, while the denominator is the area shaded in gray. The two areas coincide (i.e., degradation index is 0) when $K_{sat}$ does not degrade with decreasing C. The earlier and more pronounced the degradation, the smaller the hatched area with respect to the gray area, and therefore the degradation index will increase, with one as the highest value for immediate and complete $K_{sat}$ degradation.

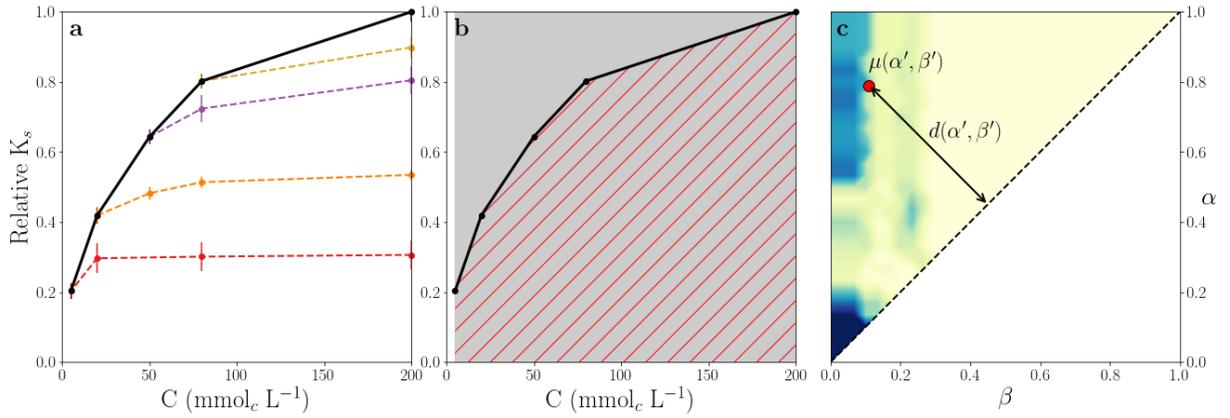

**Figure 5:** The constituents necessary for calculating the degradation and reversibility indices. Panel a shows hysteresis curves, with the degradation curve in solid black and reversal curves in dashed lines. Panel b shows the two areas necessary to calculate the degradation index. Panel c depicts the weight function $\mu(\alpha, \beta)$ corresponding to the hysteresis curves in panel a, where the distance $d(\alpha', \beta')$ of the diagonal line to given point $\mu(\alpha', \beta')$ (in red) is shown. Lighter (darker) shades indicate lower (higher) values of $\mu(\alpha, \beta)$.

The calculation of the reversibility index is more convoluted, as it is based on weight functions, $\mu(\alpha, \beta)$, first introduced in the context of irreversible $K_{sat}$ degradation by Kramer et al. (2021). We provide here a short description of the weight function and its importance, and for a full account the reader is referred to the aforementioned paper.

Any hysteresis graph, such as the one in Fig. 5a, can be fully described by a weight function, µ(α, β), that encompasses all the information of the hysteresis. The variables α and β represent the concentration of the solution, C, when it is initially decreased (β) and subsequently increased ($\alpha$). The weights in $\mu(\alpha, \beta)$ are distributed in a right triangle in the (α, β) plane, as seen if Fig. 5c. It is recommended to calculate a weight function based on a hysteresis graph with at least four or five reversal curves (Kramer et al., 2021). The closer the weights are to the diagonal line marked with a dashed line, the thinner a hysteresis loop will be. In the limit where all the weights are on the diagonal, the hysteresis loop collapses into a line, meaning that degradation is fully reversible, because the reversal curves fall exactly on the degradation curve. As weights are concentrated farther from the diagonal, the hysteresis loop gets thicker, meaning that the reversal curves will be further away from the degradation curve, exhibiting a lower degree of reversibility. Therefore, we define the reversibility index (RI) as one minus the sum of all weights in $\mu(\alpha, \beta)$, themselves weighted (multiplied) by their distance to the diagonal (*d* marked with an arrow in Fig. 5c), as follows:

$$\text{RI} = 1 - \frac{\sqrt{2}}{C_{max}} \frac{I[\mu(\alpha,\beta)\mathcal{D}(\alpha,\beta)]}{I[\mu(\alpha,\beta)]}. \qquad [5]$$

The function *I* denotes the integral over the weight triangle as follows:

$$I[\mu(\alpha, \beta)] = \int_0^{C_{max}} d\alpha \int_0^{\alpha} d\beta \, f(\alpha, \beta),$$

and

$$\mathcal{D}(\alpha, \beta) = \frac{\alpha - \beta}{\sqrt{2}}$$

is the Euclidean distance of a point $(\alpha, \beta)$ to the diagonal $\alpha = \beta$.

A reversibility index of 1 means a collapsed hysteresis loop, or total reversibility in $K_{sat}$. A reversibility index of 0 means that the hysteresis loop is as wide as possible, and there is no reversibility in $K_{sat}$ whatsoever.

With those tools in hand, we can start quantitatively comparing the soils we measured. It is important to emphasize that both the degradation and the reversibility indices are not absolute measures of a soil, but useful tools to compare the behavior of soils measured under the same experimental conditions.

Figure 6 shows the degradation index (panel a) and the reversibility index (panel b) for the three soils, as a function of SAR. As expected, soil susceptibility to degradation increases with SAR, and larger clay content is associated with greater susceptibility. Our novel reversibility experiments revealed an interesting phenomenon whereby the degree of reversibility decreases with the increase in SAR (Fig. 6b). Yet, higher clay content does not necessarily mean lesser degree of reversibility. Moreover, there is no clear relationship between a soil's degradation index and its reversibility index. For instance, at SAR 20, the sandy loam (red curve) and the sandy clay loam (blue curve) have very similar degradation indices (0.015 and 0.042, respectively), but the sandy clay loam's reversibility index is much smaller (0.253) than that of the sandy loam (1.000). In fact, the reversibility index of the sandy clay loam is even smaller than that of the clay (0.385, in black), although the clay content in the latter is much higher.

We hypothesize that susceptibility to degradation or rehabilitation does not directly translate into real risks of irreversible soil degradation. As suggested by Kramer and Mau (2020), it is the interaction between intrinsic soil properties (e.g., soil texture), management decisions (irrigation water quality and scheduling, etc) and rainfall regime that produces the actual response of the system, and therefore its associated risks. The results shown here underscore the need to further investigate $K_{sat}$ reversibility, as our understanding of the mechanisms controlling it is still insufficient.

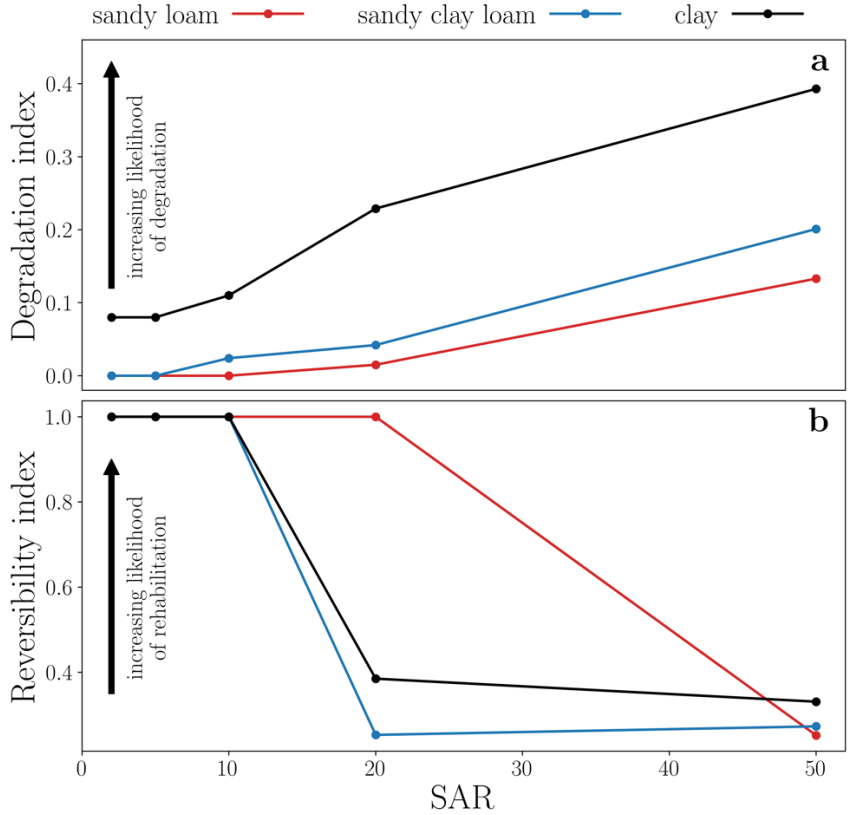

**Figure 6:** Degradation and Reversibility indices as a function of SAR for the three soils.

## 4. Discussion

The hysteresis graphs (Figs. 1-3), soil swelling (Fig. 4) and the degradation index (Fig. 6a) show a clear and known picture, namely soils are affected by sodicity, and more specifically, soils with a higher clay content are more susceptible to degradation by sodicity than coarse textured ones. More interestingly, the same hysteresis graphs reveal that $K_{sat}$ rehabilitation is highly dependent on the leaching history of the soil. The rehabilitation is soil specific and depends on the regions in the (C, SAR) parameter space, as shown by reversal curves with varying degrees of reversibility, from complete reversibility to total irreversibility.

Our working assumption is that clay swelling and dispersion are the two main processes that control changes in $K_{sat}$ following exposure to the (C, SAR) parameter space (Shainberg and Letey, 1984). Comparing the relative $K_{sat}$ curves with the swelling data may elucidate the (C, SAR) parameter space conditions under which swelling is the main cause for changes in $K_{sat}$ and those where dispersion is the main cause. For instance, in the sandy loam (10% clay), the lack of change in $K_{sat}$ for SAR 10 (Fig. 1a) and for SAR 20 up to C concentration of 10 mmol$_c$ L$^{-1}$ (Fig. 2a), is in agreement with the absence of swelling (Fig. 4). Conversely, although no swelling was noted in this soil for SAR 20, a drop in $K_{sat}$ during stage 1 of the experiment (i.e., stepwise dilution of the leaching solution) when changing from 10 to 5 mmol$_c$ L$^{-1}$ solution, followed by a complete reversibility in $K_{sat}$ during stage 2 (i.e., stepwise increase in the leaching solution concentration), was observed. This is a clear indication that clay swelling during stage 1 and subsequent shrinkage during stage 2 were the mechanisms controlling $K_{sat}$ in this case. Yet, the hysteresis in the $K_{sat}$ curves with SAR 50 solutions of 40, 20 and 15 mmol$_c$ L$^{-1}$ (Fig. 3a), is ascribed to clay dispersion, that has taken place under such high sodicity level, as no swelling was observed under these conditions (Fig. 4).

Similar argumentation can be used to explain the mechanisms that control the absence or existence of hysteresis in the $K_{sat}$ in the sandy clay loam (25% clay) and clay (40% clay) soils. Our results suggest that the weight of clay dispersion, relative to that of swelling, in determining the reversibility in $K_{sat}$ under the studied (C, SAR) parameter space, increases with increase in clay content.

Our results provide solid proof to the fact that even small variations in the parameter space may yield stark differences regarding the ability of a soil to rehabilitate. For instance, after the sandy clay loam has been leached with low concentrations of 10, 15, and 20 mmol$_c$ L$^{-1}$ (Fig. 2b), its relative $K_{sat}$ increased to 0.74, 0.91, and 1.02 (respectively), upon an increase in concentration. These observations suggest that a relatively small concentration difference of 10 mmol$_c$ L$^{-1}$ can make a difference between full rehabilitation and a 26% irreversible decrease in $K_{sat}$. A similar picture exists for the sandy loam (Fig. 3a), where degradation at concentration of 15 and 20 mmol$_c$ L$^{-1}$ could only rehabilitate to relative $K_{sat}$ of 0.58 and 0.80, respectively. It can be concluded that the impacts of swelling and dispersion on soil clays, as expressed by changes in $K_{sat}$, are very sensitive to small changes in the (C, SAR) parameter space.

As revealed by the reversal curves, there are many ways in which a soil can respond to an increase in C following an initial degradation in $K_{sat}$. We classified specific reversal curves by their degree of reversibility, and also characterized the soils using full hysteresis graphs, composed of many reversal curves. A hysteresis graph can be fully characterized by weight functions (Kramer et al., 2021), that comprehensively capture the response of a soil to any increase or decrease in soil salinity. From the weight functions and the degradation curve, we were able to derive useful metrics to quantify a soil's susceptibility to degradation (the degradation index, Fig. 6a), and the ability of a soil to rehabilitate (the reversibility index, Fig. 6b). Our results show that the degradation trends of soils associated with clay contents (and thus texture), as was specifically noted from the swelling measurements, do not determine the degree of reversibility of a soil. That, because the degree of reversibility is controlled by the magnitude of clay dispersion rather than swelling.

In principle, soils can be found in any region of the space defined by the reversibility and rehabilitation indices. Put simply, we just cannot naively assume that because a soil degrades easily, it will be resistant to rehabilitation. Nor can we say that because a soil does not degrade easily, it will be of easy rehabilitation. All conditions can exist. For example, Figure 5 shows conditions in which a soil can have high susceptibility to degradation and low rehabilitation capabilities (e.g., the clay soil at SAR 20). Let's call this a (high, low) point in the (degradation, rehabilitation) space. The clay soil at SAR 10, in turn, can be described as a (high, high) point, while the sandy clay loam at SAR 20 is a (low, low) point, and the sandy loam at SAR 20 is a (low, high) point. Therefore, measuring how a soil's history of salinity and sodicity affects its structure is therefore crucial for devising informed irrigation practices aiming at maintaining healthy soils.

## 5. Conclusions

Our results confirm both the first and the second aforementioned hypotheses, with the qualification that the effects of salinity and sodicity and the interactions between them are more complex than we first thought. Our findings indicate that soils could have any combination of high/low susceptibility to degradation and ability to rehabilitate under the (C, SAR) parameter space. The degradation and rehabilitation indices are useful tools to quantitatively compare the behavior of different soils, or even a same soil under varying environmental conditions.

In this study, we have dealt only with the case of solutions with constant SAR, and we varied the soil water concentration. Under realistic conditions, both the salinity and sodicity could change at the same time. Again, Kramer et al. (2021) developed the conceptual framework to make sense of concomitant changes in salinity and sodicity. In order to characterize the response of a soil to salinity and sodicity to a fuller extent, an extra set of experiments is warranted, that measures $K_{sat}$ for solutions of constant C and varying SAR. Although these experiments are labor intensive, we believe that the information they yield is critical to enhance our understanding of the important phenomenon of $K_{sat}$ hysteresis. We hope that highlighting the existence and relevance of hysteresis in $K_{sat}$ for sustainable use of soils, would resonate with the community, and support further studies towards promoting management practices aimed at obtaining healthier soils.

**Declaration of Competing Interest**

The authors declare that they have no known competing financial interests or personal relationships that could have appeared to influence the work reported in this paper.

**Acknowledgments**

We would like to thank Nimrod Schwartz for useful discussions.

# Supplementary Material

The following figures, S1 and S2, show the response of the three soils to solutions of SAR 2 and SAR 5, respectively. We found that for the sandy loam and sandy clay loam soils, $K_{sat}$ increases upon a decrease in C, and there is a further $K_{sat}$ increase in the reversal curves. This is a puzzling result, and we made sure that it is not the outcome of spurious artifacts. A partial explanation of this phenomenon may be attributed to the decrease in solution viscosity as concentration is decreased (Zhu et al., 2019), although the further increase in $K_{sat}$ as C is increased still requires further investigation.

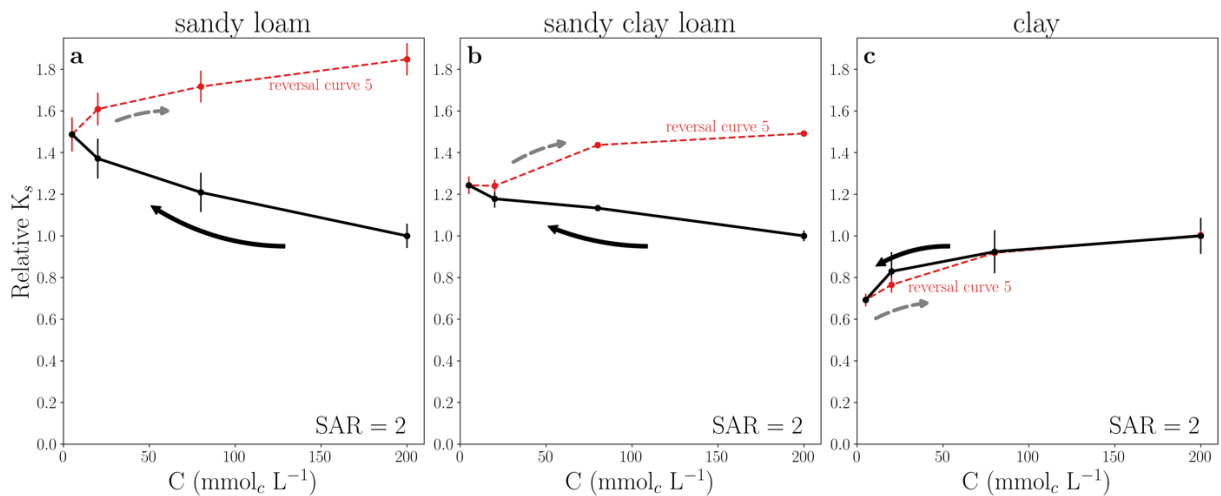

**Figure S1:** Hysteresis curves for the (a) sandy loam, (b) sandy clay loam and (c) clay soils, leached with SAR=2 solutions. Measurements start at the rightmost data point in the solid black curve and follow the arrows. Reversal curves start at the leftmost data point of the dashed lines (5 $mmol_c$ $L^{-1}$) and follow the dashed arrow. Bars indicate standard deviation.

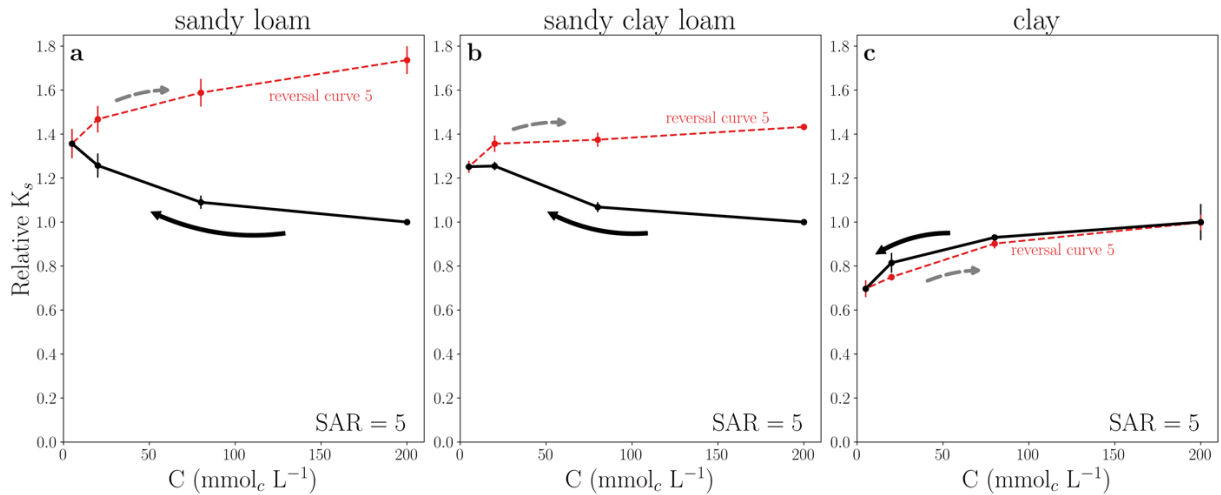

**Figure S2:** Hysteresis curves for the (a) sandy loam, (b) sandy clay loam and (c) clay soils, leached with SAR=5 solutions. Measurements start at the rightmost data point in the solid black curve and follow the arrows. Reversal curves start at the leftmost data point of the dashed lines (5 mmolc L-1) and follow the dashed arrow. Bars indicate standard deviation.

**References**

Zhu, Y., J. M. Bennett, and A. Marchuk. 2019. Reduction of hydraulic conductivity and loss of organic carbon in non-dispersive soils of different clay mineralogy is related to magnesium induced disaggregation. *Geoderma*, *349*, 1-10.